\documentclass[]{aastex631}

\submitjournal{ApJL}

\begin{document}

\title{Reconciling Parker Solar Probe observations and magnetohydrodynamic theory}

\correspondingauthor{Tommaso Alberti}
\email{tommaso.alberti@inaf.it}

\author[0000-0001-6096-0220]{Tommaso Alberti}
\affiliation{INAF-Istituto di Astrofisica e Planetologia Spaziali, 00133 Roma, Italy}

\author[0000-0002-7102-5032]{Simone Benella}
\affiliation{INAF-Istituto di Astrofisica e Planetologia Spaziali, 00133 Roma, Italy}

\author[0000-0002-3403-647X]{Giuseppe Consolini}
\affiliation{INAF-Istituto di Astrofisica e Planetologia Spaziali, 00133 Roma, Italy}

\author[0000-0002-6303-5329]{Mirko Stumpo}
\affiliation{Università degli Studi di Roma Tor Vergata, Dipartimento di Fisica, 00133 Roma, Italy}
\affiliation{INAF-Istituto di Astrofisica e Planetologia Spaziali, 00133 Roma, Italy}

\author[0000-0002-5335-7068]{Roberto Benzi}
\affiliation{Università degli Studi di Roma Tor Vergata, Dipartimento di Fisica, 00133 Roma, Italy}

\begin{abstract}
The Parker Solar Probe mission provides a unique opportunity to characterize several features of the solar wind at different heliocentric distances. Recent findings have shown a transition in the inertial range spectral and scaling properties around 0.4-0.5 au when moving away from the Sun. Here we provide, for the first time, how to reconcile these observational results on the radial evolution of the magnetic and velocity field fluctuations with two scenarios drawn from the magnetohydrodynamic theory. The observed breakdown is the result of the radial evolution of magnetic field fluctuations and plasma thermal expansion affecting the distribution between magnetic and velocity fluctuations. The two scenarios point towards an evolving nature of the coupling between fields that can be also reconciled with Kraichnan and Kolmogorov pictures of turbulence. Our findings have important implications for turbulence studies and modeling approaches.
\end{abstract}


\section{Introduction}
\label{sec:intro}

Since 2018 the Parker Solar Probe (PSP) mission is collecting solar wind plasma and magnetic field data through the inner Heliosphere, reaching the closest distance to the Sun ever reached by any previous mission \citep{Fox16,Kasper21}. Thanks to the PSP journey around the Sun (it has completed 11 orbits) a different picture has been drawn for the near-Sun solar wind with respect to the near-Earth one \citep{Kasper19,Bale19b,Malaspina20,Chhiber20,Bandyopadhyay22,Zank22}. Different near-Sun phenomena have been frequently encountered, as the emergence of magnetic field flips, i.e., the so-called switchbacks \citep{Dudok20,Zank20}, kinetic-scale current sheets \citep{Lotekar22}, and a scale-invariant population of current sheets between ion and electron inertial scales \citep{Chhiber21}. Going away from the Sun (from 0.17 au to 0.8 au), it has been shown a radial evolution of different properties of solar wind turbulence \citep{Chen20} as the spectral slope of the inertial range (moving from --3/2 close to the Sun to --5/3 at distances larger than 0.4 au), an increase of the outer scale of turbulence, a decrease of the Alfvénic flux, and a decrease of the imbalance between outward ($z^+$)- and inward ($z^-$)-propagating components \citep{Chen20}. Although the near-Sun solar wind shares different properties with the near-Earth one \citep{Allen20,Cuesta22}, significant differences have been also found in the variance of magnetic fluctuations (about two orders of magnitude) and in the compressive component of inertial range turbulence. In a similar way, \citet{Alberti20} firstly reported a breakdown of the scaling properties of the energy transfer rate, likely related to the breaking of the phase-coherence of inertial range fluctuations. These findings, also highlighted by \citet{Telloni21} and \citet{Alberti22} analyzing a radial alignment between PSP and Solar Orbiter and PSP and BepiColombo, respectively, have been interpreted as an increase in the efficiency of the nonlinear energy cascade mechanism when moving away from the Sun. More recently, by investigating the helical content of turbulence \citet{Alberti22b} highlighted a damping of magnetic helicity over the inertial range between 0.17 au and 0.6 au suggesting that the solar wind develops into turbulence by a concurrent effect of large-scale convection of helicity and creation/annihilation of helical wave structures. All these features shed new light into the radial evolution of solar wind turbulence that urge to be considered in expanding models of the solar wind \citep{Verdini19,Grappin21}, also to reproduce and investigate the role of proton heating and anisotropy of magnetic field fluctuations \citep{Hellinger15}. 

First attempts to connect observational results obtained by PSP and theoretical predictions have been mainly devoted to turbulence transport models in a nearly incompressible magnetohydrodynamic (NI MHD) framework \citep{Zank17}. As an example, \citet{Adhikari20} reported on a plausible agreement between the radial evolution of some turbulent quantities (e.g., the fluctuating kinetic energy, the correlation length) derived from PSP first orbit (between 0.17 au and 0.61 au) and numerical solutions of the NI MHD turbulence transport model \citep{Zank17}. Thus, the NI MHD model has been used to derived additional turbulent quantities as the cascade rate, the balance between inward and outward modes, and the normalized cross-helicity $\sigma_C$ and the residual energy $\sigma_R$. In agreement with previous theoretical expectations \citep{Adhikari15,Zank17,Zank18} they found that $\sigma_R$ decreases with increasing distance from the Sun, in agreement with a reasonable correlation between the most Alfvénic events ($\sigma_C \to 1$) and both increases in the energy cascade rate and local temperature \citep{Andres22}. These recent observations and theoretical findings suggest to revise an old view by \citet{Dobrowolny80} according to which an initially asymmetric MHD turbulence, as that observed by PSP close to the Sun with an abundance of outward-propagating modes $z^+$, in absence of nonlinear interactions, relaxes toward a state characterized by the absence of one of the possible modes $z^+$ or $z^-$. That is, what is the role of nonlinear interactions in generating inward-propagating ($z^-$) modes such that at larger distances from the Sun we can observe 

In this work we start with the same theoretical framework of the NI MHD proposed by \citet{Zank17} but we focus our attention on the consequences of observing an imbalanced turbulence close to the Sun, with $z^+ \gg z^-$, evolving towards a more balanced state with the radial distance, with $z^+ \sim z^-$. We find evidence of two different scenarios: an Alfvénically--dominated up to 0.3 au and a magnetically--dominated at larger distances (greater than 0.6 au). The observed breakdown is the result of the radial evolution of the distribution between magnetic and velocity fluctuations and their mutual coupling. The two scenarios can be also reconciled with Kraichnan and Kolmogorov pictures of turbulence in terms of the radial evolution of the coupling between fields. The manuscript is organized as follows: Section \ref{sec:theory} introduces the theoretical framework, while Section \ref{sec:obs} presents the observational results; finally, Section \ref{sec:conclu} summarizes the results and provides outlooks for future investigations.

\section{Theoretical background}
\label{sec:theory}

As in \citet{Zank17} we use the incompressible ($\nabla \cdot {\bf z^\pm} = 0$) MHD equations 
\begin{equation}
    \partial_t {\bf z^\pm} + \left( {\bf C_A} \cdot \nabla \right) {\bf z^\pm} + \left( {\bf z^\mp} \cdot \nabla \right) {\bf z^\pm} = -\frac{1}{\rho_0} \nabla p + \nu^\pm \nabla^2 {\bf z^\pm}, \\
    \label{eq:mhd}
\end{equation}
where ${\bf z^\pm} = {\bf v} \pm {\bf b}$ are the Elsasser variables \citep{Elsasser50}, being ${\bf v}$ the velocity field and ${\bf b} = \frac{\bf B}{\sqrt{\mu_0 \rho_0}}$ the magnetic field in Alfvén units, $C_A = \frac{B_0}{\sqrt{\mu_0 \rho_0}}$ is the background Alfvén speed, $\rho_0$ is the mass density, $p$ is the kinetic pressure, and $\nu^\pm$ are the dissipative coefficients. The Els\"asser variables describe the inward- and outward-propagating modes \citep{Elsasser50}. 

As firstly noted by \citet{Chen20} with PSP measurements outward-propagating modes $z^+$ have a stronger radial dependence with respect to inward modes $z^-$ ($z^+ \sim r^{-0.85}$ vs. $z^- \sim r^{-0.25}$), traducing into a radial trend of their ratio
\begin{equation}
    R^\pm = \frac{z^+}{z^-} \sim r^{-0.6}.
    \label{eq:ratio}
\end{equation}
This means that moving from 1 au to 0.1 au the ratio increases of a factor 4, although ${\bf z}^\pm$ show a similar spectral exponent at variance of the heliocentric distance \citep{Chen20}. Thus, close to the Sun we are in an unbalanced scenario in which $|{\bf z}^+| \gg |{\bf z}^-|$, evolving towards a balanced one $|{\bf z}^+| \sim |{\bf z}^-|$, typically observed at distances larger than 0.5-0.6 au \citep{Chen20}. Furthermore, inward-propagating modes have a longer correlation time than outward ones \citep{Chen20,Cuesta22}, thus strengthening the hypothesis that $z^-$ modes are generated via reflection of $z^+$ ones, i.e., the nonlinear term is responsible of the observed radial trend. This suggests to deeper investigate what are the consequences of Eq. (\ref{eq:ratio}). Indeed, the existence of two states, i.e., $|{\bf z}^+| \gg |{\bf z}^-|$ close to the Sun and $|{\bf z}^+| \sim |{\bf z}^-|$ at larger distances, can be traduced into a different nature of the coupling between ${\bf v}$ and ${\bf b}$ as follows. 

The condition $|{\bf z}^+| \sim |{\bf z}^-|$ means 
\begin{equation}
    |{\bf v} + {\bf b}| \sim |{\bf v} - {\bf b}|
\end{equation}
or equivalently
\begin{equation}
    |{\bf v}|^2 + |{\bf b}|^2 + 2 {\bf v} \cdot {\bf b} \sim |{\bf v}|^2 + |{\bf b}|^2 - 2 {\bf v} \cdot {\bf b}
\end{equation}
that is
\begin{equation}
    {\bf v} \cdot {\bf b} \sim 0.
\end{equation}
Thus, $|{\bf z}^+| \sim |{\bf z}^-|$ traduces into ${\bf v} \perp {\bf b}$. 

Conversely, the condition $|{\bf z}^+| \gg |{\bf z}^-|$ is
\begin{equation}
    |{\bf v} + {\bf b}| \gg |{\bf v} - {\bf b}|
\end{equation}
that corresponds to 
\begin{equation}
    |{\bf v}|^2 + |{\bf b}|^2 + 2 {\bf v} \cdot {\bf b} \gg |{\bf v}|^2 + |{\bf b}|^2 - 2 {\bf v} \cdot {\bf b},
\end{equation}
and leading to 
\begin{equation}
    {\bf v} \cdot {\bf b} \gg 0.
\end{equation}
This condition is, in principle, satisfied for any angle between the two fields $\theta_{vb} \in [0,90^\circ)$, being maximized for the case ${\bf v} \parallel {\bf b}$. Thus, the relaxation from an initially asymmetric state ($|{\bf z}^+| \gg |{\bf z}^-|$) toward a symmetric one ($|{\bf z}^+| \sim |{\bf z}^-|$) can be linked to a different nature of the ${\bf v}$--${\bf b}$ coupling, i.e., to a different degree of correlation between magnetic and velocity fields fluctuations.

As usual in MHD turbulence \citep{Matthaeus82,Roberts87,Bavassano98}, two measurable quantities can be introduced to take into account the different role of magnetic and kinetic energies as well as the relations between fields fluctuations. These two parameters are the normalized cross-helicity $\sigma_C$ and the normalized residual energy $\sigma_R$
\begin{eqnarray}
    \sigma_C &=& \frac{2 \langle {\bf v} \cdot {\bf b} \rangle}{\langle {\bf v}^2\rangle+\langle {\bf b}^2\rangle} = \frac{\langle \left({\bf z^+}\right)^2\rangle - \langle \left({\bf z^-}\right)^2\rangle}{\langle \left({\bf z^+}\right)^2\rangle+\langle \left({\bf z^-}\right)^2\rangle}, \\    \label{eq:sigmac}
    \sigma_R &=&\frac{\langle {\bf v}^2\rangle-\langle {\bf b}^2\rangle}{\langle {\bf v}^2\rangle+\langle {\bf b}^2\rangle} = \frac{R_A-1}{R_A+1},    \label{eq:sigmar}
\end{eqnarray}
with $R_A = \langle {\bf v}^2\rangle/\langle {\bf b}^2\rangle$ being the Alfvén ratio and $\langle \cdots \rangle$ stands for time average. 

\noindent$\sigma_C$ is a measure of the energy balance between outward- and inward-propagating fluctuations, while $\sigma_R$ measures the balance between kinetic and magnetic energy. $\sigma_C = \pm1$ evidences the presence of only one component ($+$: outward, $-$: inward), $|\sigma_C|<1$ corresponds to the presence of both components and/or to non-Alfvénic fluctuations, while $\sigma_R = \pm1$ evidences the existence of velocity--/magnetic--only fluctuations, with $\sigma_R = 0$ meaning equipartition. 

The two scenarios drawn above immediately give us
\begin{eqnarray}
    |{\bf z}^+| \sim |{\bf z}^-| &\Rightarrow& {\bf v} \cdot {\bf b} = 0 \Rightarrow \sigma_C = 0 \label{eq:sigma1}, \\
    |{\bf z}^+| \gg |{\bf z}^-| &\Rightarrow& \sigma_C = 1 \label{eq:sigma2}.
\end{eqnarray}
Eqs. (\ref{eq:sigma1})-(\ref{eq:sigma2}) clearly suggest that a one-to-one correspondence can be depicted with $\sigma_C$, with clear boundary values and varying between 0 and 1, while $\sigma_R$ cannot be unambiguously determined. The observed scenarios drawn in terms of the normalized cross-helicity $\sigma_C$ are in agreement with previous models, as the NI MHD model by \citet{Zank17} predicting values larger than 0 within 1 au. In the following we explore our theoretical expectations by using PSP measurements in the inner Heliosphere to compute the radial scaling of $\sigma_C$ and $\sigma_R$ as well as the joint probability of occurrence between pairs of values at different heliocentric distances. To compute $\sigma_C$ and $\sigma_R$ we use the prescriptions widely adopted in literature \citep[e.g.,][]{Bavassano98,DAmicis15}. The polarity of $z^\pm$ modes is selected to always satisfy the condition that $z^+$ is an outward-propagating fluctuation in the solar wind reference frame as seen from the Sun. Then, $\sigma_{C,R}$ are computed using running averages over a window of 1 hour length shifted by 1 minute along the full dataset. This procedure, although not preserving the independence of sub-samples \citep[as in ][]{Bavassano98}, allows us to increase the statistics. We have also verified that this preserves the significance of the results as also previously highlighted by \citet{Bruno07} using Helios 2 data.

\section{Parker Solar Probe observations}
\label{sec:obs}

\begin{figure}[h]
\plotone{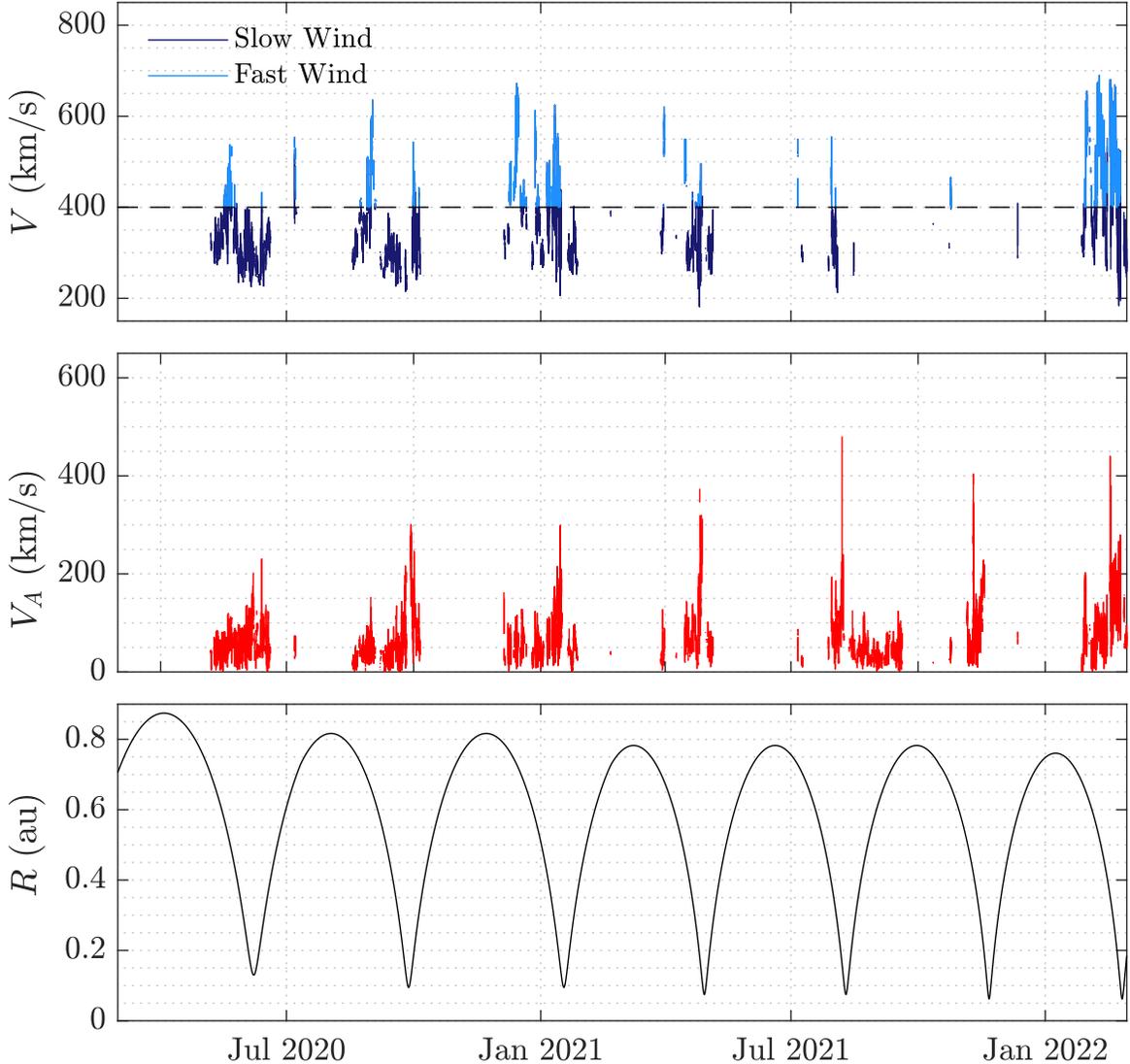}
\caption{(From top to bottom) The plasma bulk speed $V$, the Alfvén speed $V_A = 21.8 \, \frac{B}{\sqrt{n}}$ ($B$ in units of nT, $n$ in units of cm$^{-3}$), and the PSP radial distance to the Sun $R$. Dark and light blue lines in the top panel refer to slow ($V < 400$ km s$^{-1}$) and fast ($V > 400$ km s$^{-1}$) solar wind intervals, respectively.}
\label{fig1}
\end{figure}

We use PSP magnetic field and plasma measurements in the time interval from 01 March 2020 to 01 March 2022, i.e., covering four PSP perihelia corresponding to Encounters 5-11 (\url{https://sppgway.jhuapl.edu/encounters}). Specifically, magnetic field data are taken from the outboard fluxgate magnetometer (MAG) from the FIELDS instrument suite \citep{Bale16} and are L2 quality data at 1 minute time resolution, while plasma measurements are obtained by the Solar Probe Cup (SPC) of the SWEAP instrument suite \citep{Kasper16} and correspond to L3 quality data at 27.96 seconds time resolution. For our analysis, all data have been resampled at 1 minute resolution for consistency, forming a dataset of $N = 1051200$ data points and covering the heliocentric range of distances between $\sim$0.1 and $\sim$0.85 au. However, the calibrated data points ({\it good quality} to be used in the analysis) are $N_{cal} = 159375$, corresponding to rough 15\% of the full data points. Figure \ref{fig1} reports the plasma bulk speed $V$, the Alfvén speed $V_A = 21.8 \, \frac{B}{\sqrt{n}}$ ($B$ in units of nT, $n$ in units of cm$^{-3}$), and the PSP radial distance (in au) to the Sun $R$, respectively.

While a clear trend with the heliocentric distance $R$ cannot be recovered for the plasma bulk speed (as expected), a dependence on $R$ of the Alfvén speed $V_A$ seems to be present, increasing as the Sun is approached (again, as expected). This suggests that Alfvén field radially evolves according to both the large-scale configuration of the Parker spiral and the expansion of the solar wind plasma through the innermost Heliosphere as an outward-streaming gas \citep{Parker58}. 

\begin{figure}[!h]
\plotone{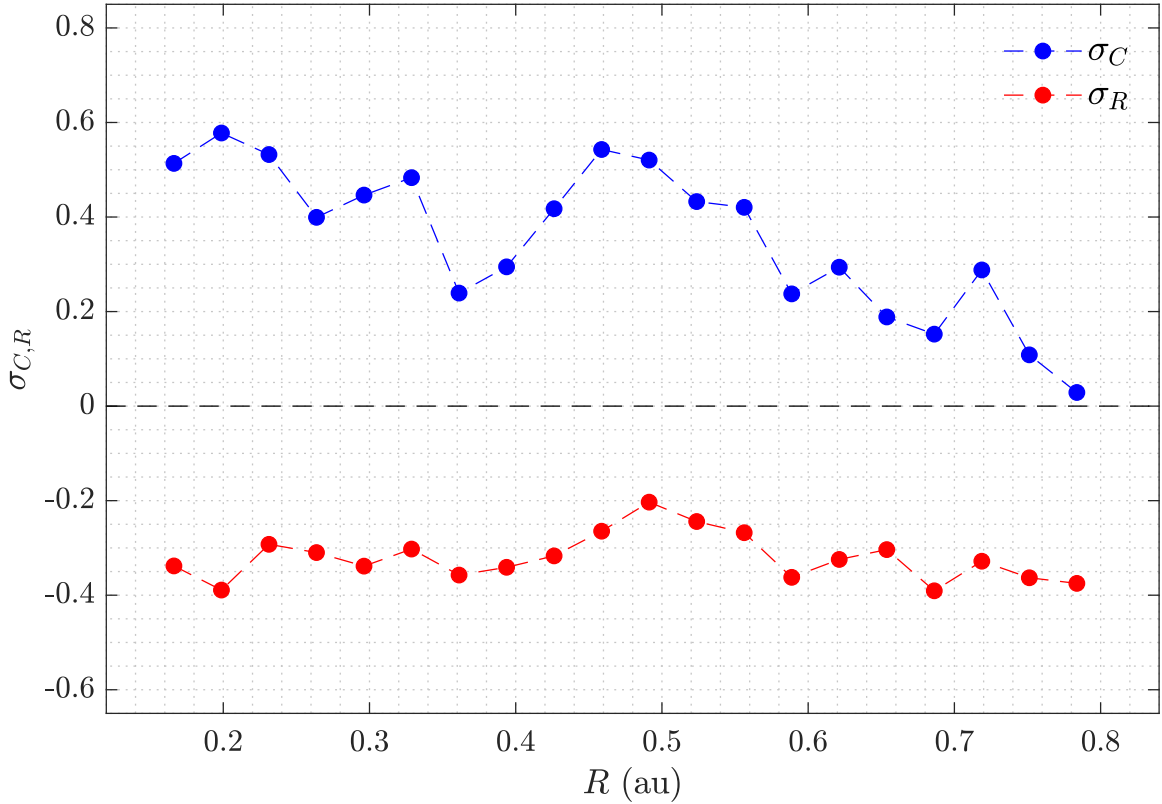}
\caption{The radial dependence of the reduced cross-helicity $\sigma_C$ (blue) and the residual energy $\sigma_R$ (red).}
\label{fig2}
\end{figure}

The first step of our analysis is to characterize the radial behavior of the reduced cross-helicity $\sigma_C$ (blue) and the residual energy $\sigma_R$ (red) as reported in Figure \ref{fig2}. 

A clear radial dependence of the normalized cross-helicity $\sigma_C$ is observed, while the residual energy $\sigma_R$ is almost constant and always negative. Our results are consistent with those recently reported by \citet{Andres22} who found a nearly constant and negative $\sigma_R$ at all heliocentric distances and an increasing $\sigma_C$ with increasing temperature (i.e., decreasing distance). Furthermore, our results are qualitatively in agreement with \citet{Adhikari20,Zank21} using the NI MHD model \citep{Zank17} who reported a decreasing $\sigma_C$ with increasing $R$, although disagreeing with the predicted behavior of $\sigma_R$ \citep[increasingly negative as $R$ increases,][]{Adhikari20}. Our results, thus, suggest a turbulent nature with prevailing 2D structures over the slab component \citep{Oughton16} as $R$ increases. Furthermore, our findings are in agreement with the two scenarios drawn in Section \ref{sec:theory} in terms of $\sigma_C$, summarized in Eqs. (\ref{eq:sigma1})-(\ref{eq:sigma2}), suggesting $\sigma_C \to 1$ close to the Sun (more precisely this condition is matched at the Alfvén point where $z^-=0$ being $v=b$) and $\sigma_C \to 0$ far away.

\begin{figure}[h!]
\plotone{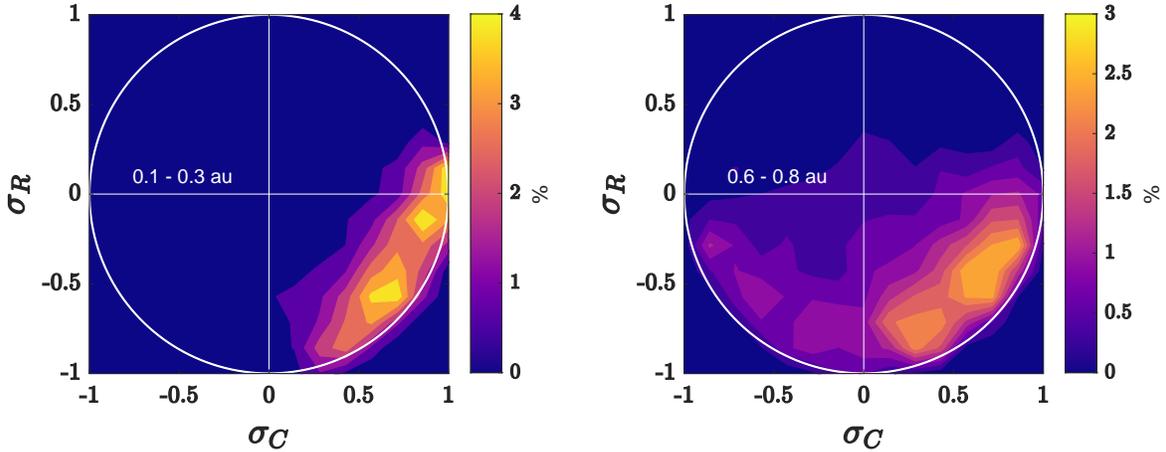}
\caption{The joint distribution of the values of the normalized cross-helicity $\sigma_C$ and the normalized residual energy $\sigma_R$ within two heliocentric ranges: close to the Sun ($R \in [0.1,0.3]$ au) and far away ($R \in [0.6-0.8]$ au).}
\label{fig3}
\end{figure}

To further exploit the nature of these two scenarios we evaluate the joint distribution of the values of the cross-helicity $\sigma_C$ and the residual energy $\sigma_R$ within two different bands of heliocentric distances: close to the Sun ($R \in [0.1,0.3]$ au) and far away ($R \in [0.6-0.8]$ au). The choice of these two heliocentric ranges is consistent with previous observations reporting a different nature of the turbulent properties, changing around 0.4-0.5 au \citep{Chen20,Alberti20,Stumpo21}. The results are shown in Figure \ref{fig3}.

Moving away from the Sun the distribution of pairs of $\left(\sigma_C, \sigma_R\right)$ evolve from the fourth quadrant $\left(\sigma_C>0, \sigma_R<0\right)$ at 0.1-0.3 au towards the third one $\left(\sigma_C<0, \sigma_R<0\right)$ at 0.6-0.8 au. On one side, the decreases in the cross-helicity $\sigma_C$ suggests an evolution from a more to a less Alfvénic turbulence, while the observed trend for the residual energy $\sigma_R$ suggests a turbulence always dominated by magnetic fluctuations. Furthermore, although the predominant modes are always outward-propagating fluctuations ($z^+$), the presence of a non-null probability of observing $\sigma_C < 0$ at 0.6-0.8 au implies an increase in the occurrence of inward fluctuations ($\sigma_C < 0$). This suggests that as we move away from the Sun the nonlinear term is becoming relevant, being able to generate inward-propagating modes. Interestingly, this is almost independent on $\sigma_R$, i.e., in terms of energy of fluctuations there is no evidence of a dynamical transition from magnetic to kinetic predominance. We return on this point in Section \ref{sec:conclu}. Thus, our findings are in agreement with the simple theoretical framework introduced in Eqs. (\ref{eq:sigma1})-(\ref{eq:sigma2}), as well as with recently published literature \citep{Andres22}, and describe a transition between two states that can be then classified as:

\begin{enumerate}
    \item[(i)] $\sigma_C > 0, \sigma_R<0$ for $R < 0.3$ au: this corresponds to a magnetically--dominated state with the predominance of outward Alfvénic fluctuations;
    \item[(ii)] $\sigma_C \in [-1,1], \sigma_R < 0$ for $R > 0.6$ au: this corresponds to the predominance of a magnetically--dominated state with non-Alfvénic fluctuations or with the presence of both outward and inward modes.
\end{enumerate}

\begin{figure}[h!]
\plotone{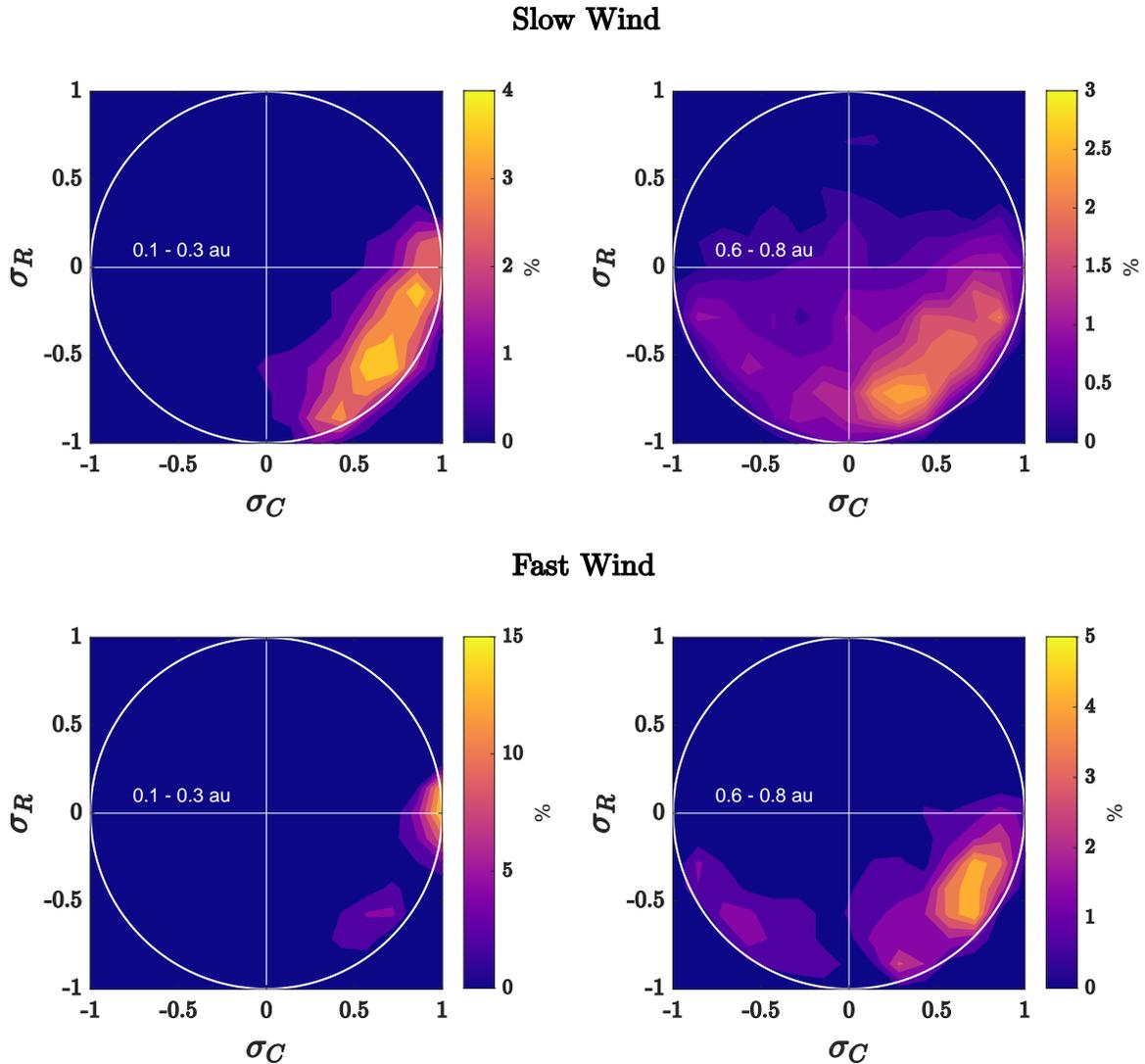}
\caption{As in Figure \ref{fig3} but separating slow ($V<400$ km/s, upper panels) and fast ($V>400$ km/s, lower panels) solar wind intervals.}
\label{fig4}
\end{figure}

As a final step of our analysis, since the cross-helicity $\sigma_C$ also depends on the solar wind speed, we investigated the joint distribution of pairs $\left(\sigma_C, \sigma_R\right)$ within the same two heliocentric ranges ($R \in [0.1,0.3]$ au and $R \in [0.6-0.8]$ au) by separating slow ($V<400$ km/s) and fast ($V>400$ km/s) solar wind intervals (Figure \ref{fig1}, top panel). The results are reported in Figure \ref{fig4}. 

The results look interesting since we can highlight a clear different role in terms of solar wind streams. In particular, we observe a trend from $\left(\sigma_C, \sigma_R\right) = \left(1,0\right)$ to $\sigma_R < 0$ for the fast solar wind and a radial decreasing in terms of $\sigma_R$ for the slow solar wind. The observed behavior of $\sigma_C$ is again in agreement with our theoretical predictions, i.e., Eqs. (\ref{eq:sigma1})-(\ref{eq:sigma2}), but also with previous models as the NI MHD one \citep{Zank17,Adhikari20} and/or energy transfer rate estimations \citep{Andres22}. Furthermore, differently from the overall features of the solar wind (see Figure \ref{fig2}), i.e., when considering together fast and slow streams, the observed decreasing $\sigma_R$ with increasing $R$ for the slow solar wind is in agreement with the NI MHD model \citep{Adhikari20}. This suggests that the NI MHD model can be particularly useful for investigating the radial evolution of solar wind turbulent quantities for slow solar wind streams. Thus, the energy-containing range for the slow solar wind can be described as a superposition of a (predominant) 2D component and a (less dominant) slab one \citep{Zank21}, matching our findings when separating fast and slow winds.

\section{Conclusions}
\label{sec:conclu}

As a final task we discuss implications of our findings, trying to interpret them in the framework of turbulence. Earlier studies \citep[e.g.,][]{Chen20,Alberti20} using PSP observations have shown that an MHD Alfvénic scenario is reached when approaching the Sun for the spectral and the scaling properties of the Els\"asser field fluctuations, although mainly dominated by one mode (specifically, $z^+$), as well as for both the magnetic and the velocity field fluctuations across the inertial range, with a spectral exponent close to --3/2 \citep{Chen20}. Conversely, at distances larger than 0.6 au all fields are characterized by a spectral exponent close to --5/3 \citep{Chen20}, and both modes are almost equi-partitioned ($|{\bf z^+}|/|{\bf z^-}|\sim 1$). According to the earlier work by \citet{Dobrowolny80} an initially asymmetric MHD turbulence $|{\bf z}^+| \gg |{\bf z}^-|$, like that observed close to the Sun by PSP \citep{Chen20}, in absence of nonlinear interactions, should relax toward a state characterized by the presence of only one of the possible modes ${\bf z}^+$ or ${\bf z}^-$. Our main result is that the final state is not characterized by the absence of one of the two Alfvénic modes but that we are observing a different nature of the ${\bf v}$--${\bf b}$ coupling (Eqs. (\ref{eq:sigma1})-(\ref{eq:sigma2})), linked to the more/less Alfvénic nature of the solar wind close/far-away from the Sun. This can also explain why close to the Sun an MHD Alfvénic turbulence {\em à la} Kraichnan is observed \citep{Kraichnan65}, with a spectral exponent --3/2, while close to the Earth a fluid turbulence scenario {\em à la} Kolmogorov \citep{Kolmogorov41}, with $\beta = -5/3$, can be drawn \citep{Chen20,Alberti20}. Our findings are also in agreement with models of balanced turbulence \citep{Goldreich95,Schekochihin20}, suggesting that the observed changes in the spectral exponent can be related to a relaxation of the balanced turbulence scenario \citep{Chen20}. 

Since the {\it missing} element in the theoretical framework proposed by \citet{Dobrowolny80} is the nonlinear term, we now discuss the fundamental implications of our results, in terms of the ${\bf v}$--${\bf b}$ coupling, on the term $\left({\bf z^\mp} \cdot \nabla\right) {\bf z^\pm}$, both for modelling approaches and for observational results. The nonlinear term (NL) can be written as
\begin{equation}
    \text{NL} = \left({\bf z^+} \cdot \nabla \right) {\bf z^-} = \left[ \left({\bf v}+{\bf b}\right) \cdot \nabla\right] \left({\bf v}-{\bf b}\right),
\end{equation}
that is
\begin{equation}
    \text{NL} = \left[\left({\bf v} \cdot \nabla\right) {\bf v} - \left({\bf b} \cdot \nabla\right) {\bf b} \right] + \left[\left({\bf b} \cdot \nabla\right) {\bf v} - \left({\bf v} \cdot \nabla\right) {\bf b} \right].
    \label{eq:NL}
\end{equation}
The first term on the right-side of Eq. (\ref{eq:NL}) is the energetic part of the nonlinear term, i.e., it is related to the difference of the kinetic and the magnetic energy density. The second term on the right-side of Eq. (\ref{eq:NL}) is the mutual relation between ${\bf v}$ and ${\bf b}$. Indeed, due to the incompressible nature of Eqs. (\ref{eq:mhd}) the second term on the right-side of Eq. (\ref{eq:NL}) can be written as
\begin{equation}
    \left[\left({\bf b} \cdot \nabla\right) {\bf v} - \left({\bf v} \cdot \nabla\right) {\bf b} \right] = \nabla \times \left({\bf v} \times {\bf b}\right).
    \label{eq:sigma3}
\end{equation}
This allows to revisit Eqs. (\ref{eq:sigma1})-(\ref{eq:sigma2}) as
\begin{eqnarray}
    |{\bf z}^+| \sim |{\bf z}^-| &\Rightarrow& {\bf v} \cdot {\bf b} = 0 \Rightarrow \sigma_C = 0 \Rightarrow \left| \nabla \times \left({\bf v} \times {\bf b}\right) \right| \ne 0  \label{eq:sigma1n}, \\
    |{\bf z}^+| \gg |{\bf z}^-| &\Rightarrow& \sigma_C = 1 \Rightarrow \left| \nabla \times \left({\bf v} \times {\bf b}\right) \right| \simeq 0 \label{eq:sigma2n}.
\end{eqnarray}
Thus, moving away from the Sun an additional term appears in the nonlinear term that can be responsible of the observed radial behavior of some turbulence quantities, as $\sigma_C$ but also the spectral/scaling properties, being related to the nature of the ${\bf v}-{\bf b}$ coupling. Thus, more efforts are needed to describe the evolution of the helical component of turbulence in the inner Heliosphere that cannot be interpreted in a simple transport--like scenario but needs to be properly framed out in an evolving scenario, also involving the role of field coupling and intermittency \citep{Schekochihin20}. 

Our results needs to be further assessed with more and more PSP orbits as well as with observations of the sub-Alfvénic region that could open a completely different framework for the early stages of the solar wind turbulence evolution when leaving the Sun \citep{Kasper21}. A critical view of the role of the turbulent cascade in the solar wind is needed, searching for novel models of the solar wind expansion that could be at the basis of the observed scenarios. Indeed, it has been recently demonstrated how including the expansion in solar wind modeling allows to observe nearly equal spectral exponents for the Els\"asser fields, as observed, also reproducing the observed variability of spectral indices at larger distances \citep{Verdini19,Grappin21}.

\section*{Acknowledgements}

We acknowledge the NASA Parker Solar Probe Mission, the SWEAP team led by J. Kasper and the FIELDS team led by S. D. Bale for use of data. The data can be downloaded from the NASA CDAWeb (https://cdaweb.gsfc. nasa.gov/pub/data/psp/). M.S. acknowledges the PhD course in Astronomy, Astrophysics and Space Science of the University of Rome “Sapienza”, University of Rome “Tor Vergata” and Italian National Institute for Astrophysics (INAF), Italy.

\bibliography{Albertietal.bib}{}

\begin{thebibliography}{}
\expandafter\ifx\csname natexlab\endcsname\relax\def\natexlab#1{#1}\fi
\providecommand{\url}[1]{\href{#1}{#1}}
\providecommand{\dodoi}[1]{doi:~\href{http://doi.org/#1}{\nolinkurl{#1}}}
\providecommand{\doeprint}[1]{\href{http://ascl.net/#1}{\nolinkurl{http://ascl.net/#1}}}
\providecommand{\doarXiv}[1]{\href{https://arxiv.org/abs/#1}{\nolinkurl{https://arxiv.org/abs/#1}}}

\bibitem[{{Adhikari} {et~al.}(2015){Adhikari}, {Zank}, {Bruno}, {Telloni},
  {Hunana}, {Dosch}, {Marino}, \& {Hu}}]{Adhikari15}
{Adhikari}, L., {Zank}, G.~P., {Bruno}, R., {et~al.} 2015, \apj, 805, 63,
  \dodoi{10.1088/0004-637X/805/1/63}

\bibitem[{{Adhikari} {et~al.}(2020){Adhikari}, {Zank}, {Zhao}, {Kasper},
  {Korreck}, {Stevens}, {Case}, {Whittlesey}, {Larson}, {Livi}, \&
  {Klein}}]{Adhikari20}
{Adhikari}, L., {Zank}, G.~P., {Zhao}, L.~L., {et~al.} 2020, \apjs, 246, 38,
  \dodoi{10.3847/1538-4365/ab5852}

\bibitem[{Alberti {et~al.}(2020)Alberti, Laurenza, Consolini, Milillo,
  Marcucci, Carbone, \& Bale}]{Alberti20}
Alberti, T., Laurenza, M., Consolini, G., {et~al.} 2020, The Astrophysical
  Journal, 902, 84, \dodoi{10.3847/1538-4357/abb3d2}

\bibitem[{{Alberti} {et~al.}(2022){Alberti}, {Milillo}, {Heyner}, {Hadid},
  {Auster}, {Richter}, \& {Narita}}]{Alberti22}
{Alberti}, T., {Milillo}, A., {Heyner}, D., {et~al.} 2022, The Astrophys. J.,
  926, 174, \dodoi{10.3847/1538-4357/ac478d}

\bibitem[{Alberti {et~al.}(2022)Alberti, Narita, Hadid, Heyner, Milillo,
  Plainaki, Auster, \& Richter}]{Alberti22b}
Alberti, T., Narita, Y., Hadid, L.~Z., {et~al.} 2022, A\&A, 664, L8,
  \dodoi{10.1051/0004-6361/202244314}

\bibitem[{Allen {et~al.}(2020)Allen, Lario, Odstrcil, Ho, Jian, Cohen, Badman,
  Jones, Arge, Mays, Mason, Bale, Bonnell, Case, Christian, de~Wit, Goetz,
  Harvey, Henney, Hill, Kasper, Korreck, Larson, Livi, MacDowall, Malaspina,
  McComas, McNutt, Mitchell, Pulupa, Raouafi, Schwadron, Stevens, Whittlesey,
  \& Wiedenbeck}]{Allen20}
Allen, R.~C., Lario, D., Odstrcil, D., {et~al.} 2020, The Astrophysical Journal
  Supplement Series, 246, 36, \dodoi{10.3847/1538-4365/ab578f}

\bibitem[{{Andr{\'e}s} {et~al.}(2022){Andr{\'e}s}, {Sahraoui}, {Huang},
  {Hadid}, \& {Galtier}}]{Andres22}
{Andr{\'e}s}, N., {Sahraoui}, F., {Huang}, S., {Hadid}, L.~Z., \& {Galtier}, S.
  2022, \aap, 661, A116, \dodoi{10.1051/0004-6361/202142994}

\bibitem[{Bale {et~al.}(2016)Bale, Goetz, Harvey, Turin, Bonnell, De~Wit,
  Ergun, MacDowall, Pulupa, Andr{\'e}, {et~al.}}]{Bale16}
Bale, S., Goetz, K., Harvey, P., {et~al.} 2016, Space Science Reviews, 204, 49

\bibitem[{{Bale} {et~al.}(2019){Bale}, {Badman}, {Bonnell}, {Bowen}, {Burgess},
  {Case}, {Cattell}, {Chandran}, {Chaston}, {Chen}, {Drake}, {de Wit},
  {Eastwood}, {Ergun}, {Farrell}, {Fong}, {Goetz}, {Goldstein}, {Goodrich},
  {Harvey}, {Horbury}, {Howes}, {Kasper}, {Kellogg}, {Klimchuk}, {Korreck},
  {Krasnoselskikh}, {Krucker}, {Laker}, {Larson}, {MacDowall}, {Maksimovic},
  {Malaspina}, {Martinez-Oliveros}, {McComas}, {Meyer-Vernet}, {Moncuquet},
  {Mozer}, {Phan}, {Pulupa}, {Raouafi}, {Salem}, {Stansby}, {Stevens}, {Szabo},
  {Velli}, {Woolley}, \& {Wygant}}]{Bale19b}
{Bale}, S.~D., {Badman}, S.~T., {Bonnell}, J.~W., {et~al.} 2019, Nature, 576,
  237, \dodoi{10.1038/s41586-019-1818-7}

\bibitem[{Bandyopadhyay {et~al.}(2022)Bandyopadhyay, Matthaeus, McComas,
  Chhiber, Usmanov, Huang, Livi, Larson, Kasper, Case, Stevens, Whittlesey,
  Romeo, Bale, Bonnell, de~Wit, Goetz, Harvey, MacDowall, Malaspina, \&
  Pulupa}]{Bandyopadhyay22}
Bandyopadhyay, R., Matthaeus, W.~H., McComas, D.~J., {et~al.} 2022, The
  Astrophysical Journal Letters, 926, L1, \dodoi{10.3847/2041-8213/ac4a5c}

\bibitem[{{Bavassano} {et~al.}(1998){Bavassano}, {Pietropaolo}, \&
  {Bruno}}]{Bavassano98}
{Bavassano}, B., {Pietropaolo}, E., \& {Bruno}, R. 1998, \jgr, 103, 6521,
  \dodoi{10.1029/97JA03029}

\bibitem[{{Bruno} {et~al.}(2007){Bruno}, {D'Amicis}, {Bavassano}, {Carbone}, \&
  {Sorriso-Valvo}}]{Bruno07}
{Bruno}, R., {D'Amicis}, R., {Bavassano}, B., {Carbone}, V., \&
  {Sorriso-Valvo}, L. 2007, Annales Geophysicae, 25, 1913,
  \dodoi{10.5194/angeo-25-1913-2007}

\bibitem[{Chen {et~al.}(2020)Chen, Bale, Bonnell, Borovikov, Bowen, Burgess,
  Case, Chandran, De~Wit, Goetz, {et~al.}}]{Chen20}
Chen, C., Bale, S., Bonnell, J., {et~al.} 2020, The Astrophysical Journal
  Supplement Series, 246, 53

\bibitem[{Chhiber {et~al.}(2021)Chhiber, Matthaeus, Bowen, \& Bale}]{Chhiber21}
Chhiber, R., Matthaeus, W.~H., Bowen, T.~A., \& Bale, S.~D. 2021, The
  Astrophysical Journal Letters, 911, L7

\bibitem[{Chhiber {et~al.}(2020)Chhiber, Goldstein, Maruca, Chasapis,
  Matthaeus, Ruffolo, Bandyopadhyay, Parashar, Qudsi, de~Wit, Bale, Bonnell,
  Goetz, Harvey, MacDowall, Malaspina, Pulupa, Kasper, Korreck, Case, Stevens,
  Whittlesey, Larson, Livi, Velli, \& Raouafi}]{Chhiber20}
Chhiber, R., Goldstein, M.~L., Maruca, B.~A., {et~al.} 2020, The Astrophysical
  Journal Supplement Series, 246, 31, \dodoi{10.3847/1538-4365/ab53d2}

\bibitem[{{Cuesta} {et~al.}(2022){Cuesta}, {Parashar}, {Chhiber}, \&
  {Matthaeus}}]{Cuesta22}
{Cuesta}, M.~E., {Parashar}, T.~N., {Chhiber}, R., \& {Matthaeus}, W.~H. 2022,
  arXiv e-prints, arXiv:2202.01874.
\newblock \doarXiv{2202.01874}

\bibitem[{{D'Amicis} \& {Bruno}(2015)}]{DAmicis15}
{D'Amicis}, R., \& {Bruno}, R. 2015, \apj, 805, 84,
  \dodoi{10.1088/0004-637X/805/1/84}

\bibitem[{de~Wit {et~al.}(2020)de~Wit, Krasnoselskikh, Bale, Bonnell, Bowen,
  Chen, Froment, Goetz, Harvey, Jagarlamudi, {et~al.}}]{Dudok20}
de~Wit, T.~D., Krasnoselskikh, V.~V., Bale, S.~D., {et~al.} 2020, The
  Astrophysical Journal Supplement Series, 246, 39,
  \dodoi{10.3847/1538-4365/ab5853}

\bibitem[{{Dobrowolny} {et~al.}(1980){Dobrowolny}, {Mangeney}, \&
  {Veltri}}]{Dobrowolny80}
{Dobrowolny}, M., {Mangeney}, A., \& {Veltri}, P. 1980, Phys. Rev. Lett., 45,
  144, \dodoi{10.1103/PhysRevLett.45.144}

\bibitem[{{Elsasser}(1950)}]{Elsasser50}
{Elsasser}, W.~M. 1950, Physical Review, 79, 183,
  \dodoi{10.1103/PhysRev.79.183}

\bibitem[{Fox {et~al.}(2016)Fox, Velli, Bale, Decker, Driesman, Howard, Kasper,
  Kinnison, Kusterer, Lario, {et~al.}}]{Fox16}
Fox, N., Velli, M., Bale, S., {et~al.} 2016, Space Science Reviews, 204, 7,
  \dodoi{10.1007/s11214-015-0211-6}

\bibitem[{{Goldreich} \& {Sridhar}(1995)}]{Goldreich95}
{Goldreich}, P., \& {Sridhar}, S. 1995, \apj, 438, 763, \dodoi{10.1086/175121}

\bibitem[{{Grappin} {et~al.}(2021){Grappin}, {Verdini}, \&
  {M{\"u}ller}}]{Grappin21}
{Grappin}, R., {Verdini}, A., \& {M{\"u}ller}, W.~C. 2021, in SF2A-2021:
  Proceedings of the Annual meeting of the French Society of Astronomy and
  Astrophysics. Eds.: A. Siebert, ed. A.~{Siebert}, K.~{Bailli{\'e}},
  E.~{Lagadec}, N.~{Lagarde}, J.~{Malzac}, J.~B. {Marquette}, M.~{N'Diaye},
  J.~{Richard}, \& O.~{Venot}, 222--225

\bibitem[{{Hellinger} {et~al.}(2015){Hellinger}, {Matteini}, {Landi},
  {Verdini}, {Franci}, \& {Tr{\'a}vn{\'\i}{\v{c}}ek}}]{Hellinger15}
{Hellinger}, P., {Matteini}, L., {Landi}, S., {et~al.} 2015, The Astrophys. J.
  Lett., 811, L32, \dodoi{10.1088/2041-8205/811/2/L32}

\bibitem[{{Kasper} {et~al.}(2016){Kasper}, {Abiad}, {Austin}, {Balat-Pichelin},
  {Bale}, {Belcher}, {Berg}, {Bergner}, {Berthomier}, {Bookbinder}, {Brodu},
  {Caldwell}, {Case}, {Chandran}, {Cheimets}, {Cirtain}, {Cranmer}, {Curtis},
  {Daigneau}, {Dalton}, {Dasgupta}, {DeTomaso}, {Diaz-Aguado}, {Djordjevic},
  {Donaskowski}, {Effinger}, {Florinski}, {Fox}, {Freeman}, {Gallagher},
  {Gary}, {Gauron}, {Gates}, {Goldstein}, {Golub}, {Gordon}, {Gurnee}, {Guth},
  {Halekas}, {Hatch}, {Heerikuisen}, {Ho}, {Hu}, {Johnson}, {Jordan},
  {Korreck}, {Larson}, {Lazarus}, {Li}, {Livi}, {Ludlam}, {Maksimovic},
  {McFadden}, {Marchant}, {Maruca}, {McComas}, {Messina}, {Mercer}, {Park},
  {Peddie}, {Pogorelov}, {Reinhart}, {Richardson}, {Robinson}, {Rosen},
  {Skoug}, {Slagle}, {Steinberg}, {Stevens}, {Szabo}, {Taylor}, {Tiu}, {Turin},
  {Velli}, {Webb}, {Whittlesey}, {Wright}, {Wu}, \& {Zank}}]{Kasper16}
{Kasper}, J.~C., {Abiad}, R., {Austin}, G., {et~al.} 2016, \ssr, 204, 131,
  \dodoi{10.1007/s11214-015-0206-3}

\bibitem[{{Kasper} {et~al.}(2019){Kasper}, {Bale}, {Belcher}, {Berthomier},
  {Case}, {Chandran}, {Curtis}, {Gallagher}, {Gary}, {Golub}, {Halekas}, {Ho},
  {Horbury}, {Hu}, {Huang}, {Klein}, {Korreck}, {Larson}, {Livi}, {Maruca},
  {Lavraud}, {Louarn}, {Maksimovic}, {Martinovic}, {McGinnis}, {Pogorelov},
  {Richardson}, {Skoug}, {Steinberg}, {Stevens}, {Szabo}, {Velli},
  {Whittlesey}, {Wright}, {Zank}, {MacDowall}, {McComas}, {McNutt}, {Pulupa},
  {Raouafi}, \& {Schwadron}}]{Kasper19}
{Kasper}, J.~C., {Bale}, S.~D., {Belcher}, J.~W., {et~al.} 2019, Nature, 576,
  228, \dodoi{10.1038/s41586-019-1813-z}

\bibitem[{Kasper {et~al.}(2021)Kasper, Klein, Lichko, Huang, Chen, Badman,
  Bonnell, Whittlesey, Livi, Larson, Pulupa, Rahmati, Stansby, Korreck,
  Stevens, Case, Bale, Maksimovic, Moncuquet, Goetz, Halekas, Malaspina,
  Raouafi, Szabo, MacDowall, Velli, Dudok~de Wit, \& Zank}]{Kasper21}
Kasper, J.~C., Klein, K.~G., Lichko, E., {et~al.} 2021, Phys. Rev. Lett., 127,
  255101, \dodoi{10.1103/PhysRevLett.127.255101}

\bibitem[{{Kolmogorov}(1941)}]{Kolmogorov41}
{Kolmogorov}, A. 1941, Akademiia Nauk SSSR Doklady, 30, 301

\bibitem[{{Kraichnan}(1965)}]{Kraichnan65}
{Kraichnan}, R.~H. 1965, Physics of Fluids, 8, 1385, \dodoi{10.1063/1.1761412}

\bibitem[{{Lotekar} {et~al.}(2022){Lotekar}, {Vasko}, {Phan}, {Bale}, {Bowen},
  {Halekas}, {Artemyev}, {Khotyaintsev}, \& {Mozer}}]{Lotekar22}
{Lotekar}, A., {Vasko}, I.~Y., {Phan}, T., {et~al.} 2022, arXiv e-prints,
  arXiv:2202.12341.
\newblock \doarXiv{2202.12341}

\bibitem[{{Malaspina} {et~al.}(2020){Malaspina}, {Halekas},
  {Ber{\v{c}}i{\v{c}}}, {Larson}, {Whittlesey}, {Bale}, {Bonnell}, {Dudok de
  Wit}, {Ergun}, {Howes}, {Goetz}, {Goodrich}, {Harvey}, {MacDowall}, {Pulupa},
  {Case}, {Kasper}, {Korreck}, {Livi}, \& {Stevens}}]{Malaspina20}
{Malaspina}, D.~M., {Halekas}, J., {Ber{\v{c}}i{\v{c}}}, L., {et~al.} 2020, The
  Astrophys. J. Suppl. Ser., 246, 21, \dodoi{10.3847/1538-4365/ab4c3b}

\bibitem[{{Matthaeus} \& {Goldstein}(1982)}]{Matthaeus82}
{Matthaeus}, W.~H., \& {Goldstein}, M.~L. 1982, \jgr, 87, 6011,
  \dodoi{10.1029/JA087iA08p06011}

\bibitem[{{Oughton} {et~al.}(2016){Oughton}, {Matthaeus}, {Wan}, \&
  {Parashar}}]{Oughton16}
{Oughton}, S., {Matthaeus}, W.~H., {Wan}, M., \& {Parashar}, T. 2016, Journal
  of Geophysical Research (Space Physics), 121, 5041,
  \dodoi{10.1002/2016JA022496}

\bibitem[{{Parker}(1958)}]{Parker58}
{Parker}, E.~N. 1958, The Astrophys. J., 128, 664, \dodoi{10.1086/146579}

\bibitem[{{Roberts} {et~al.}(1987){Roberts}, {Klein}, {Goldstein}, \&
  {Matthaeus}}]{Roberts87}
{Roberts}, D.~A., {Klein}, L.~W., {Goldstein}, M.~L., \& {Matthaeus}, W.~H.
  1987, \jgr, 92, 11021, \dodoi{10.1029/JA092iA10p11021}

\bibitem[{{Schekochihin}(2020)}]{Schekochihin20}
{Schekochihin}, A.~A. 2020, arXiv e-prints, arXiv:2010.00699.
\newblock \doarXiv{2010.00699}

\bibitem[{{Stumpo} {et~al.}(2021){Stumpo}, {Quattrociocchi}, {Benella},
  {Alberti}, \& {Consolini}}]{Stumpo21}
{Stumpo}, M., {Quattrociocchi}, V., {Benella}, S., {Alberti}, T., \&
  {Consolini}, G. 2021, Atmosphere, 12, 321, \dodoi{10.3390/atmos12030321}

\bibitem[{Telloni {et~al.}(2021)Telloni, Sorriso-Valvo, Woodham, Panasenco,
  Velli, Carbone, Zank, Bruno, Perrone, Nakanotani, Shi, D'Amicis, Marco,
  Jagarlamudi, Steinvall, Marino, Adhikari, Zhao, Liang, Tenerani, Laker,
  Horbury, Bale, Pulupa, Malaspina, MacDowall, Goetz, de~Wit, Harvey, Kasper,
  Korreck, Larson, Case, Stevens, Whittlesey, Livi, Owen, Livi, Louarn,
  Antonucci, Romoli, O'Brien, Evans, \& Angelini}]{Telloni21}
Telloni, D., Sorriso-Valvo, L., Woodham, L.~D., {et~al.} 2021, The
  Astrophysical Journal Letters, 912, L21, \dodoi{10.3847/2041-8213/abf7d1}

\bibitem[{{Verdini} {et~al.}(2019){Verdini}, {Grappin}, {Montagud-Camps},
  {Landi}, {Franci}, \& {Papini}}]{Verdini19}
{Verdini}, A., {Grappin}, R., {Montagud-Camps}, V., {et~al.} 2019, Nuovo
  Cimento C Geophysics Space Physics C, 42, 17,
  \dodoi{10.1393/ncc/i2019-19017-x}

\bibitem[{{Zank} {et~al.}(2017){Zank}, {Adhikari}, {Hunana}, {Shiota}, {Bruno},
  \& {Telloni}}]{Zank17}
{Zank}, G.~P., {Adhikari}, L., {Hunana}, P., {et~al.} 2017, \apj, 835, 147,
  \dodoi{10.3847/1538-4357/835/2/147}

\bibitem[{{Zank} {et~al.}(2018){Zank}, {Adhikari}, {Hunana}, {Tiwari}, {Moore},
  {Shiota}, {Bruno}, \& {Telloni}}]{Zank18}
---. 2018, \apj, 854, 32, \dodoi{10.3847/1538-4357/aaa763}

\bibitem[{Zank {et~al.}(2020)Zank, Nakanotani, Zhao, Adhikari, \&
  Kasper}]{Zank20}
Zank, G.~P., Nakanotani, M., Zhao, L.-L., Adhikari, L., \& Kasper, J. 2020, The
  Astrophysical Journal, 903, 1, \dodoi{10.3847/1538-4357/abb828}

\bibitem[{{Zank} {et~al.}(2021){Zank}, {Zhao}, {Adhikari}, {Telloni}, {Kasper},
  \& {Bale}}]{Zank21}
{Zank}, G.~P., {Zhao}, L.~L., {Adhikari}, L., {et~al.} 2021, Physics of
  Plasmas, 28, 080501, \dodoi{10.1063/5.0055692}

\bibitem[{Zank {et~al.}(2022)Zank, Zhao, Adhikari, Telloni, Kasper, Stevens,
  Rahmati, \& Bale}]{Zank22}
Zank, G.~P., Zhao, L.-L., Adhikari, L., {et~al.} 2022, The Astrophysical
  Journal Letters, 926, L16, \dodoi{10.3847/2041-8213/ac51da}

\end{thebibliography}
\bibliographystyle{aasjournal}

\end{document}